\documentclass{webofc}
\usepackage[varg]{txfonts}   % Web of Conferences font
\newcommand{\gevc}{\mathrm{GeV}/c}

\newcommand{\pt}{p_{\rm T}}
\newcommand{\DtoKpi}{{\rm D^0\to K^-\pi^+}}
\newcommand{\DtoKpipi}{{\rm D^+\to K^-\pi^+\pi^+}}
\newcommand{\DstartoDpi}{{\rm D^{*+}\to D^0\pi^+}}
\newcommand{\Dzero}{{\rm D^0}}
\newcommand{\Dstar}{{\rm D^{*+}}}
\newcommand{\Dplus}{{\rm D^+}}
\newcommand{\Ds}{{\rm D_s^+}}
\newcommand{\Dstophipi}{{\rm D_s^+\to \phi\pi^+}}
\newcommand{\Lc}        {\mbox{$\mathrm {\Lambda_{c}}$}\xspace}
\newcommand{\Lb}{\ensuremath{\rm {\Lambda_b}}\xspace}
\newcommand{\Xic}        {\mbox{$\mathrm {\Xi_{c}^{0}}$}\xspace}
\newcommand{\Xib}        {\mbox{$\mathrm {\Xi_{b}^{0}}$}\xspace}

\newcommand{\Lcplus}{\ensuremath{\rm {\Lambda_c^+}\xspace}}

\newcommand{\LctopKpi}{\ensuremath{\rm \Lambda_{c}^{+}\to p K^-\pi^+}\xspace}

\newcommand{\LctopKzS}{\ensuremath{\rm \Lambda_{c}^{+}\to p K^{0}_{S}}\xspace}
\newcommand{\LctoenuLambda}{\ensuremath{\rm \Lambda_{c}^{+}\to e^{+} \nu_{e} \Lambda}\xspace}
\newcommand{\XictoenuXi}{\ensuremath{\rm \Xi_{c}^{0}\to e^{+} \nu_{e} \Xi^{-}}\xspace}

\newcommand{\pp}{\mbox{pp} }
\newcommand{\pPb}{\mbox{p--Pb} }

\newcommand {\sqrtSnn}   {\ensuremath{\sqrt{s_{\textsc{nn}}}}}
\newcommand {\sqrtS}     {\ensuremath{\sqrt{s}}}

\newcommand {\Rppb}       {\ensuremath{R_\mathrm{pPb}}}

\newcommand {\Qcp}       {\ensuremath{Q_\mathrm{CP}}}

\newcommand {\mass}     {\mbox{\rm MeV$\kern-0.15em /\kern-0.12em c^2$}}
\newcommand {\tev}      {\mbox{${\rm TeV}$}}

\newcommand{\Jpsi}      {\mbox{J\kern-0.02em /\kern-0.05em$\psi$}}

\begin{document}
\title{Charmed meson and baryon measurements in \pp and \pPb collisions with ALICE at the LHC}

\author{\firstname{Jaime Norman}\inst{1}
	%\inst{2}
	\fnsep\thanks{\email{jaime.norman@cern.ch}} \lastname{for the ALICE collaboration}
%        \firstname{Jaime} \lastname{Norman}\inst{2}\fnsep\thanks{\email{Mail address for second
%             author if necessary}} \and
%        \firstname{Third author} \lastname{Third author}\inst{3}\fnsep\thanks{\email{Mail address for last
%             author if necessary}}
        % etc.
}

\institute{The University of Liverpool, Liverpool L69 7ZE, United Kingdom 
%\and
%           Laboratoire de Physique Subatomique \& Cosmologie,53 Avenue des Martyrs, 38000 Grenoble, France
%\and
%           Last address
          }

\abstract{%
   We present here recent open heavy-flavour results from the ALICE experiment, including measurements of D-meson, $\Lc$ baryon and $\mathrm{\Xi_c^0}$ baryon production in pp collisions at $\sqrtS = 7~\tev$ and $\pPb$ collisions at $\sqrtSnn = 5.02~\tev$.
}
\maketitle

\section{Introduction}
\label{intro}

The measurement of charm production in pp collisions is an important test of perturbative QCD, and in p--Pb collisions the study of charm production can help disentangle cold nuclear matter effects from the effects modifying the $p_\mathrm{T}$ spectrum of charmed hadrons in Pb--Pb collisions due to the high-temperature and high energy-density medium formed.
The relative abundance of baryons and mesons can shed light on the process of fragmentation - a non-perturbative process - and deviations from measurements made at $e^+e^-$ colliders may hint at specific processes occuring in the higher partonic density environment of pp and $\pPb$ collisions. In addition, these measurements also provide a reference for future measurements in nucleus-nucleus collisions, where the baryon-to-meson ratio is expected to be sensitive to modified hadronisation mechanisms such as coalescence \cite{Oh:2009zj}. Preliminary results from STAR \cite{Xie:2017jcq} hint at an enhanced $\Lcplus / \Dzero$ ratio in Au--Au collisions.

\section{Heavy-flavour reconstruction at ALICE}
\label{reconstruction}

Heavy-flavour decays are reconstructed at ALICE in the central barrel, which consists of the Inner Tracking System (ITS), the Time Projection Chamber (TPC) and the Time-of-Flight detector (TOF), covering the entire azimuthal range and designed to track and identify charged particles over a wide momentum range. 
%A full description of ALICE's subdetectors can be found in \cite{Aamodt:2008zz}. 
Charmed hadrons are reconstructed at mid-rapidity ($\left|y\right| < 0.8$) via their hadronic decays including $\DtoKpi$, $\DtoKpipi$, $\DstartoDpi$, $\Dstophipi$, $\LctopKpi$ and $\LctopKzS$. Selection is made on the hadron decay topology, the signal is extracted via an invariant mass analysis, and corrections are made for the efficiency, acceptance, and the fraction of non-prompt hadrons in the signal sample. The semileptonic decays of $\Lcplus$ and $\Xic$ baryons are also reconstructed via the decay channels $\LctoenuLambda$ and $(\Xib \rightarrow )\XictoenuXi$. Here the analyses are not based on topological selections, and are instead based on subtracting the `wrong-sign' $e^-\Lambda$($e^-\Xi^-$) pair spectra from the `right-sign' $e^+\Lambda$($e^+\Xi^-$) spectra. Additional corrections include correcting for contributions to the wrong-sign spectra from $\Lb$($\Xib$), and unfolding the $e^-\Lambda$($e^-\Xi^-$) $\pt$ spectra to obtain the $\Lcplus$($\Xic$) $\pt$ spectrum. A correction for feed-down from $\Xi_c^{+,0}$ is also included for the $\Lcplus$ measurement.

\section{Results}

\begin{figure}[h]
\centering
\includegraphics[width=5cm,clip]{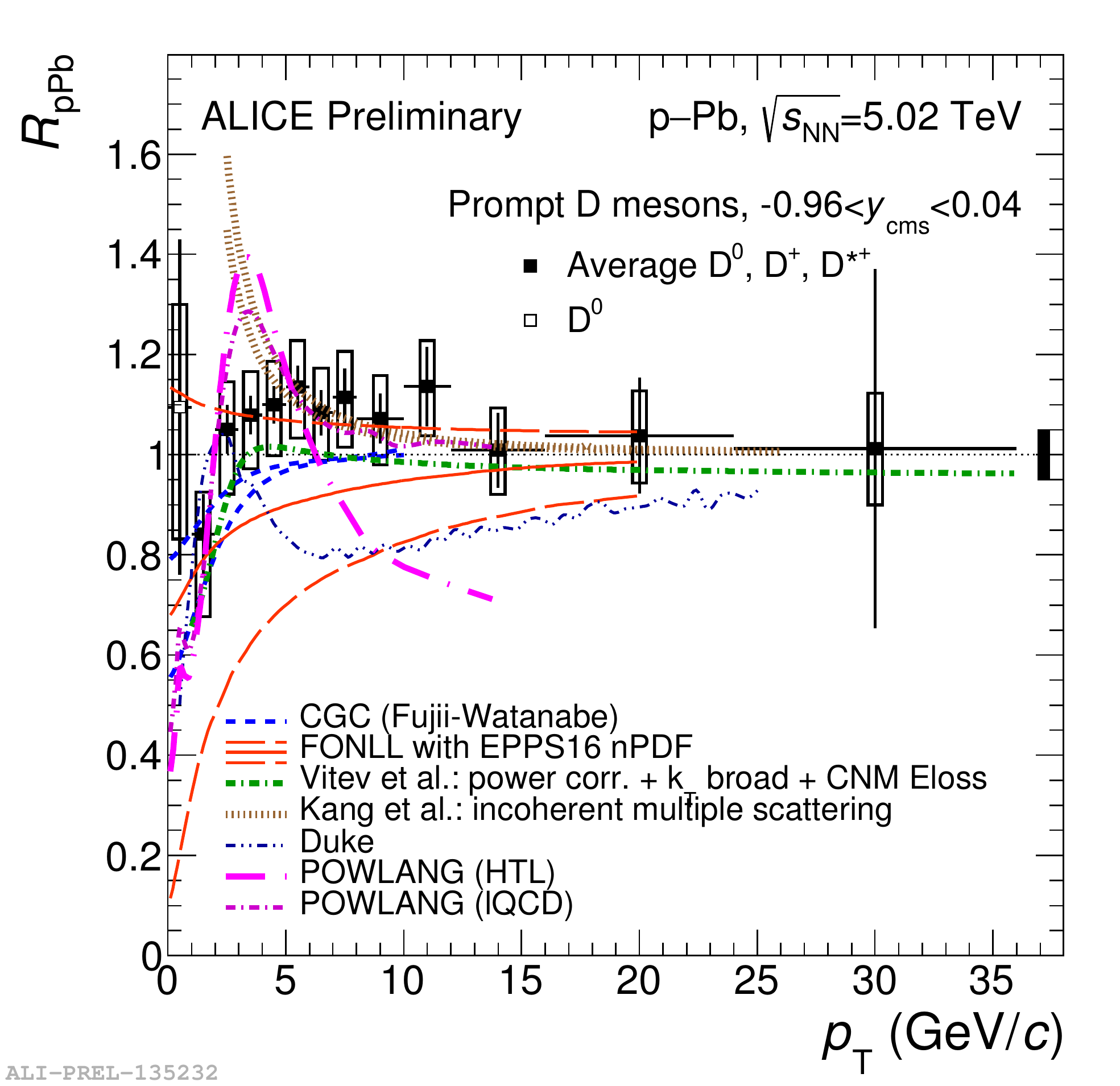}
\includegraphics[width=5cm,clip]{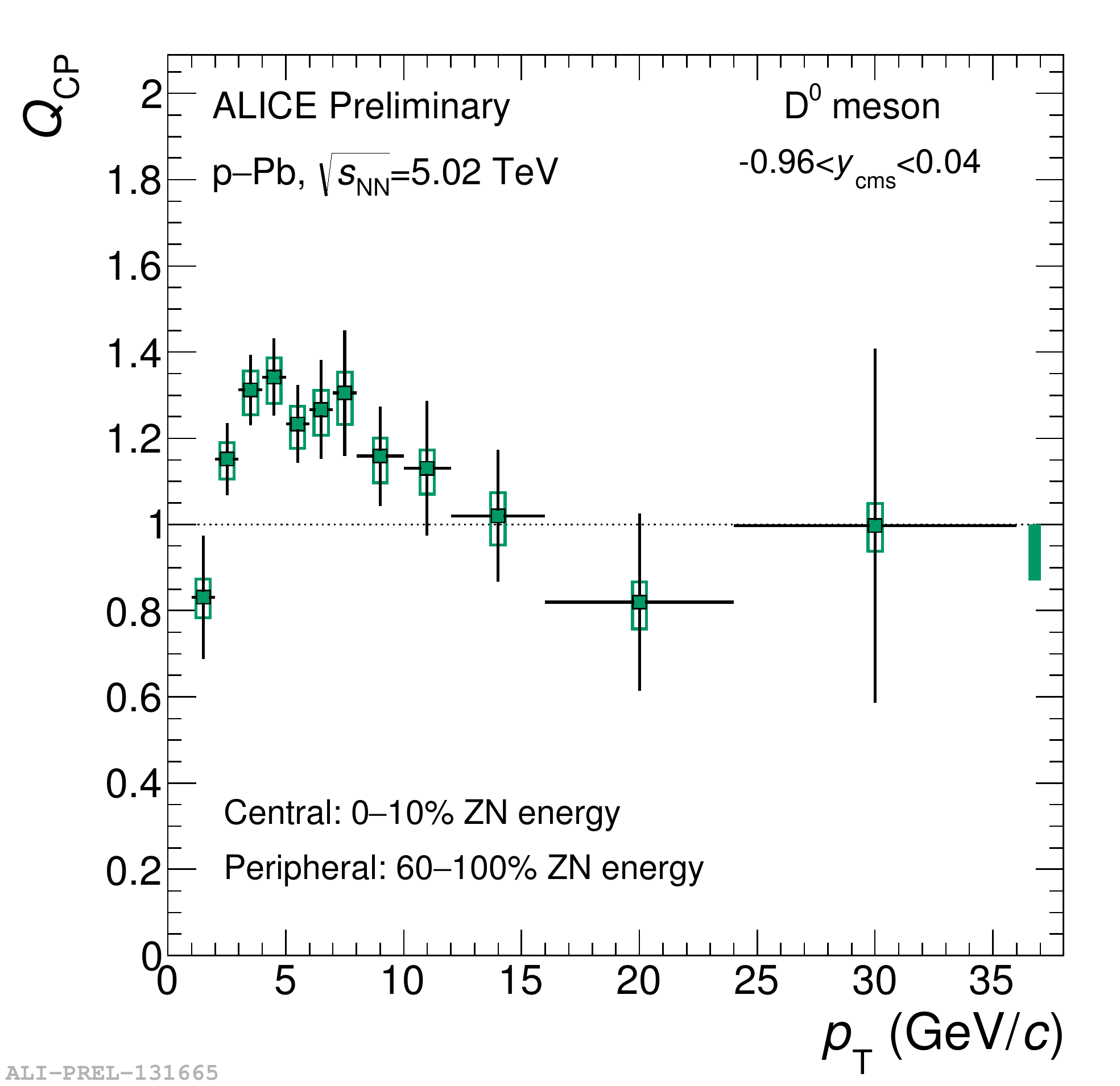}
\caption{Left: averaged $\Dzero$, $\Dplus$ and $\Dstar$ $\Rppb$ in $\pPb$ collisons at $\sqrtSnn = 5.02 ~\tev$ compared to model calculations. Right: prompt $\Dzero$ $\Qcp$ in $\pPb$ collisions at $\sqrtSnn = 5.02 ~\tev$.}
\label{fig:Dmeson-rppb}       % Give a unique label
\end{figure}

The nuclear modification factor ($\Rppb$) of $\Dzero$, $\Dplus$, $\Dstar$ and $\Ds$ mesons was measured in $\pPb$ collisions at $\sqrtSnn = 5.02 ~\tev$ \cite{ALICE-PUBLIC-2017-008}. 
%The reference for this measurement was measured in pp collisions at $\sqrtS = 7 ~\tev$ and scaled to $\sqrtS = 5.02 ~\tev$ using D-meson production predictions at both energies.
Figure \ref{fig:Dmeson-rppb} (left) shows the $\pt$-differential averaged $\Dzero$, $\Dplus$ and $\Dstar$ $\Rppb$ in comparison with models that include cold nuclear matter effects 
%\cite{MANGANO1992295,Eskola:2016oht,Sharma:2009hn,Fujii:2013yja,Kang:2014hha} 
and models that assume a Quark-Gluon Plasma is formed and include hydrodynamical effects 
%\cite{Xu:2015iha,Beraudo:2015wsd}.
\cite{ALICE-PUBLIC-2017-008}.
The statistical precision of the measurements has been improved by approximately a factor of 2 with respect to the previous measurement \cite{Adam:2016ich} due to an increased integrated luminosity. The models describe the data well, although a suppression larger than 15-20\% for $\pt>5 ~\gevc$, expected from the POWLANG(HTL) and Duke models, is slightly disfavoured by the data.
D-meson production has also been measured as a function of centrality in $\pPb$ collisions at $\sqrtSnn = 5.02 ~\tev$ \cite{ALICE-PUBLIC-2017-008}. Figure \ref{fig:Dmeson-rppb} (right) shows the $\Dzero$ $\Qcp$ calculated as the ratio of the $\Dzero$-meson nuclear modification factor in central (0-10\%) and peripheral (60-100\%) centrality intervals. 
%The centrality is measured based on energy deposited in the ZDC by slow nucleons emitted in the nuclear fragmentation process. 
%The $\Qppb$ is consistent with unity for both centrality intervals. 
The $\Qcp$ tends to increases in the interval $1 < \pt < 4 ~\gevc$ and reaches about 1.25, and then decreases in the interval $7 < \pt< 24 ~\gevc$. The average value of the $\Dzero$ $\Qcp$ is larger than unity in the interval $3 < \pt < 8 ~\gevc$ by 1.7 standard deviations of the statistical and systematic uncertainty.

\begin{figure}[h]
	\centering
	\includegraphics[width=5cm,clip]{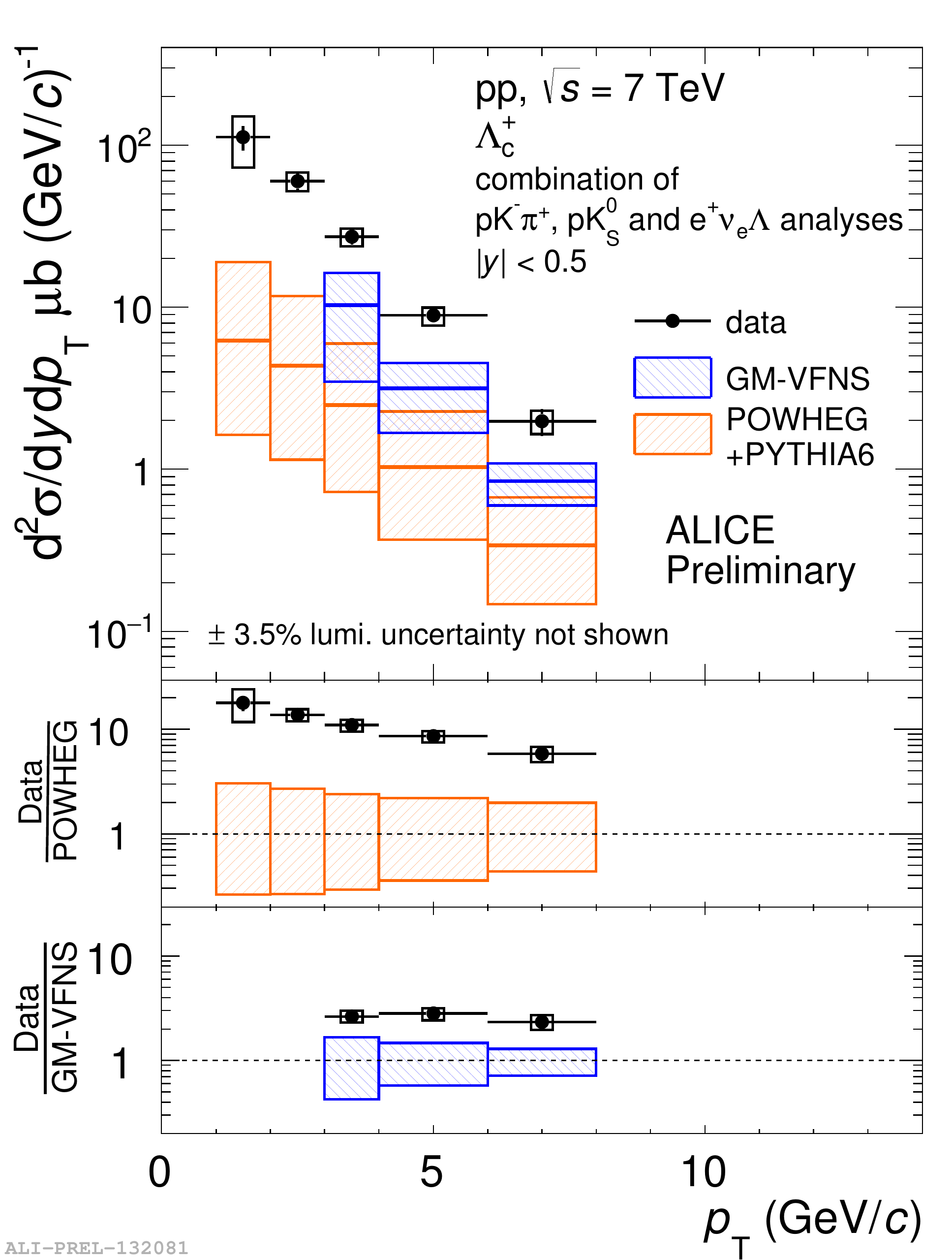}
	\includegraphics[width=5cm,clip]{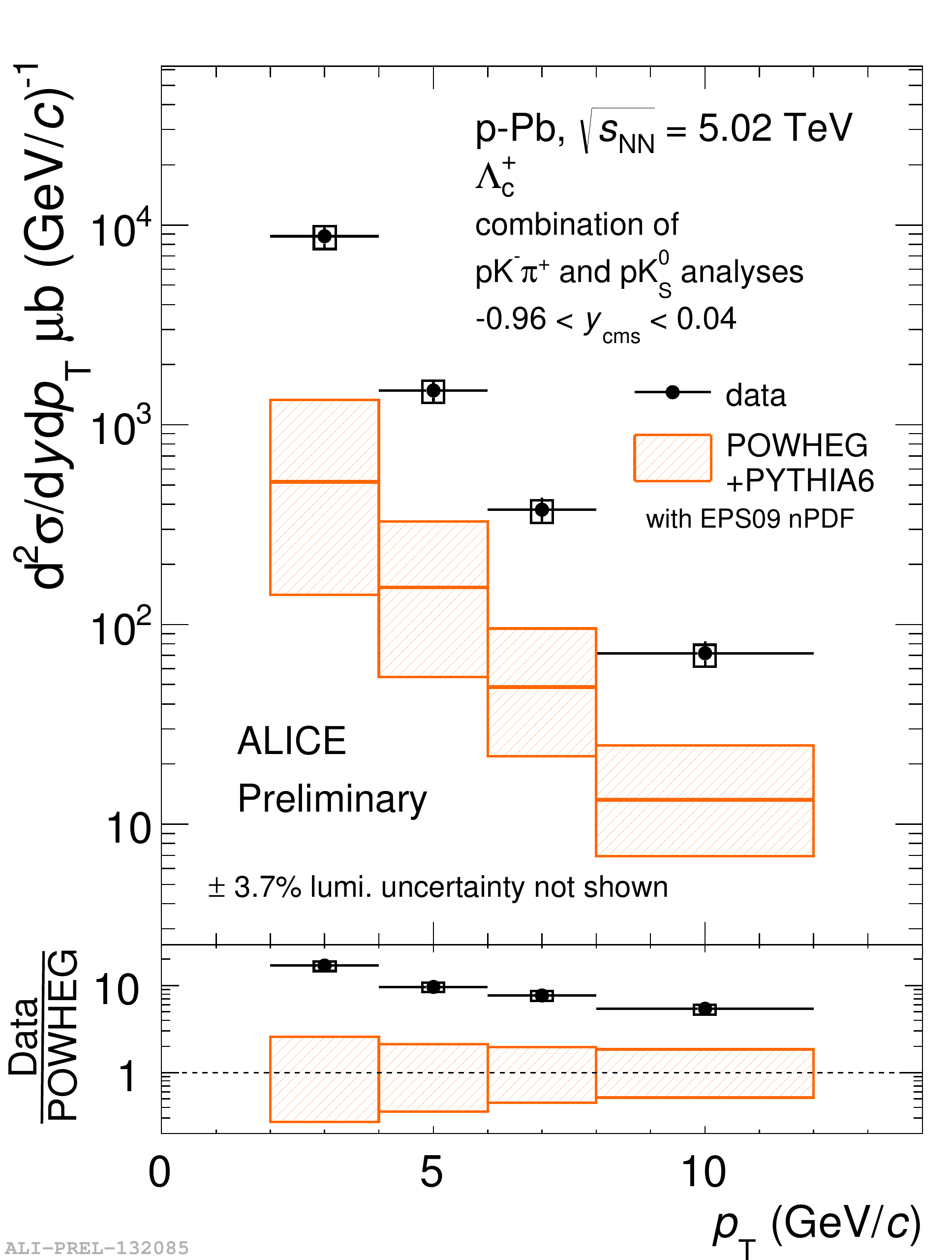}
	\caption{Left: the $\pt$-differential cross section of prompt $\Lcplus$ in pp collisions at $\sqrtS = 7 ~\tev$. Right: the $\pt$-differential cross section of prompt $\Lcplus$ in $\pPb$ collisions at $\sqrtSnn = 5.02 ~\tev$. Both measurements are compared with theoretical predictions.}
	\label{fig:Lc-cross}       % Give a unique label
\end{figure}

\begin{figure}[h]
	\centering
	\includegraphics[width=5.5cm,clip]{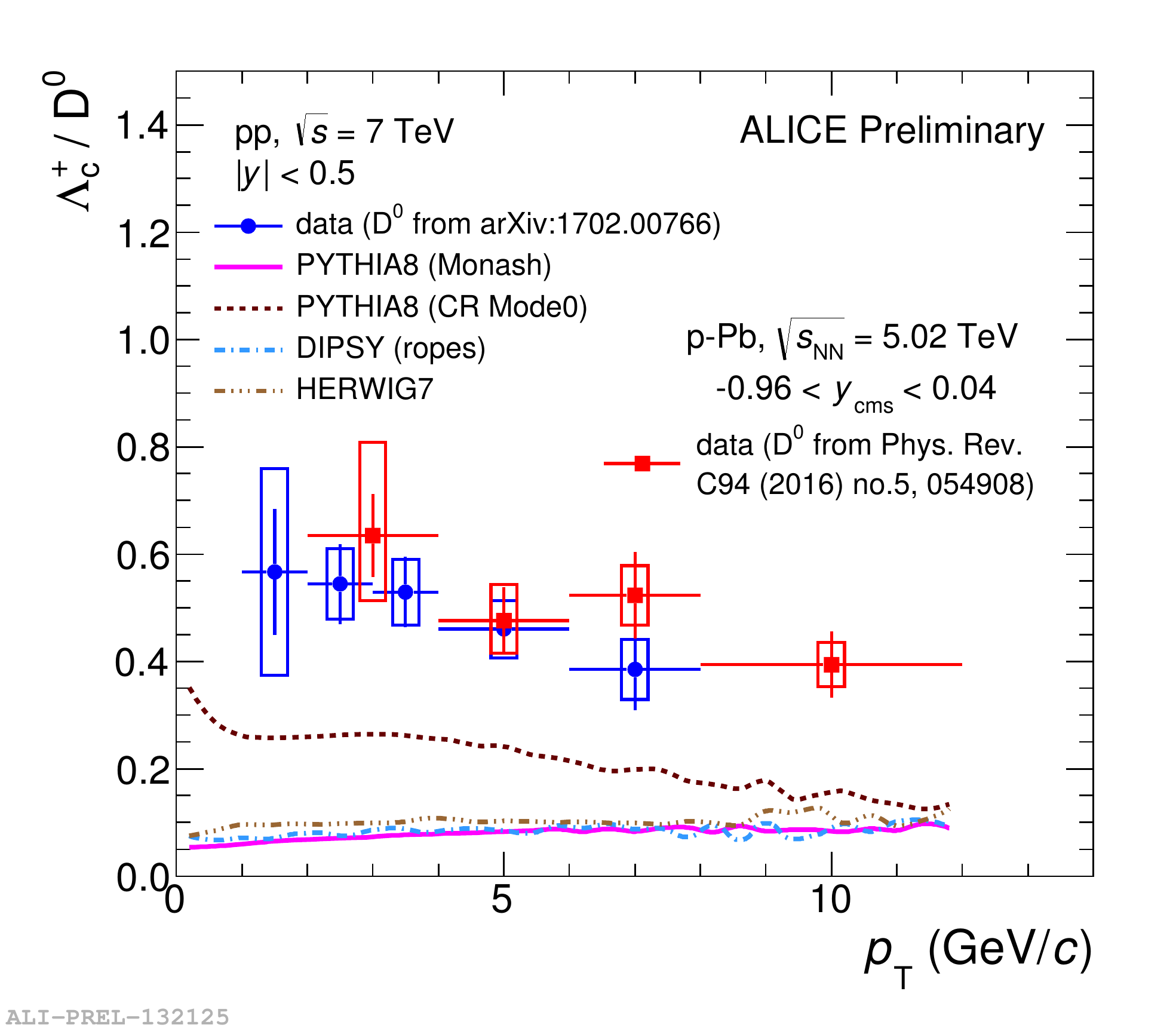}
	\includegraphics[width=8.5cm,clip]{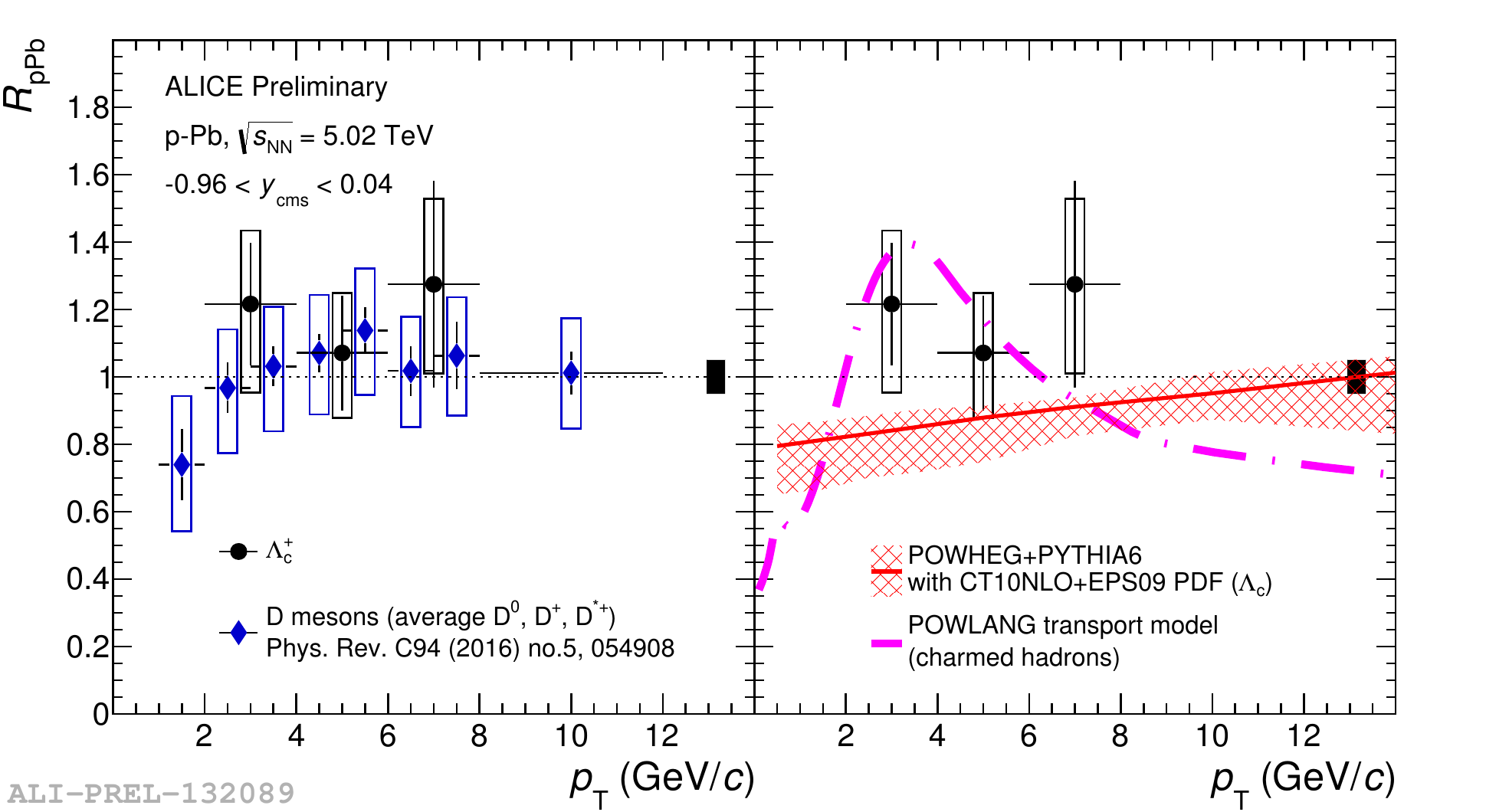}
	\caption{Left: the $\Lcplus / \Dzero$ ratio in $\pp$ collisions at $\sqrtS = 7 ~\tev$, and in $\pPb$ collisions at $\sqrtSnn = 5.02 ~\tev$, compared to model predictions. Right: the nuclear modification factor $\Rppb$ of $\Lcplus$ compared to the $\Rppb$ of D mesons, and compared with model predictions.}
	\label{fig:Lc-ratios}       % Give a unique label
\end{figure}

%The charmed baryon $\Lcplus$ has been measured using a combination
The $\pt$-differential cross section of the $\Lcplus$ baryon was measured in pp collisions at $\sqrtS = 7 ~\tev$ and in $\pPb$ collisions at $\sqrtSnn = 5.02 ~\tev$, and is reported in figure \ref{fig:Lc-cross}. The cross section is compared to perturbative calculations at NLO using the GM-VFNS \cite{Kniehl:2005mk,Kniehl:2012ti} scheme in pp collisions, and at NLO with {\sc powheg} \cite{Frixione:2007nw} matched with {\sc pythia \small6.4.25} \cite{Sjostrand:2006za}
%for parton showers and hadronisation 
in $\pp$ and $\pPb$ collisions. For the $\pPb$ {\sc powheg} the EPS09 nuclear PDF was used \cite{Eskola:2009uj}. GM-VFNS underpredicts the data by a factor of 2.5 on average, and {\sc powheg} underpredicts the data by a factor of 18(4) at low(high) $\pt$ in \pp collisions, and by a similar amount in \pPb collisions.
%as reported in the left and right panels of figure \ref{Lc-cross} respectively.
%
Figure \ref{fig:Lc-ratios} (left) shows the baryon-to-meson ratio $\Lcplus/\Dzero$ measured using the $\Lcplus$ cross sections presented in this paper and the $\Dzero$ cross sections measured by ALICE \cite{Acharya:2017jgo,Adam:2016ich}. In the same figure theoretical predictions in $\pp$ collisions are shown including {\sc pythia\small8} with and without a tune including enhanced colour reconnection \cite{Christiansen:2015yqa}, {\sc dipsy} \cite{Flensburg:2011kk} with rope hadronisation and {\sc herwig\small7} \cite{Bahr:2008pv} with hadronisation via clusters. 
The $\Lcplus / \Dzero$ ratio in pp collisions is compatible with the same ratio in $\pPb$ collisions within uncertainties.
While all models underpredict the data, {\sc pythia\small8} with enhanced colour reconnection brings the prediction closer to data.
Figure \ref{fig:Lc-ratios} (right) shows the nuclear modification factor $\Rppb$ for the $\Lcplus$ baryon in $\pPb$ collisions at $\sqrtSnn = 5.02 ~\tev$, in comparison to the averaged D-meson ($\Dzero$, $\Dplus$, $\Dstar$) $\Rppb$ \cite{Adam:2016ich}, and predictions including {\sc powheg+pythia} with EPS09 nuclear PDF \cite{Eskola:2009uj} and a prediction for charmed hadrons which assumes a small-size QGP is formed \cite{Beraudo:2015wsd}. The $\Rppb$ of $\Lcplus$ is consistent with unity and with the D-meson $\Rppb$, and does not allow to distinguish between the models presented within the current experimental uncertainties.

\begin{figure}[h]
	\centering
	\includegraphics[width=6cm,clip]{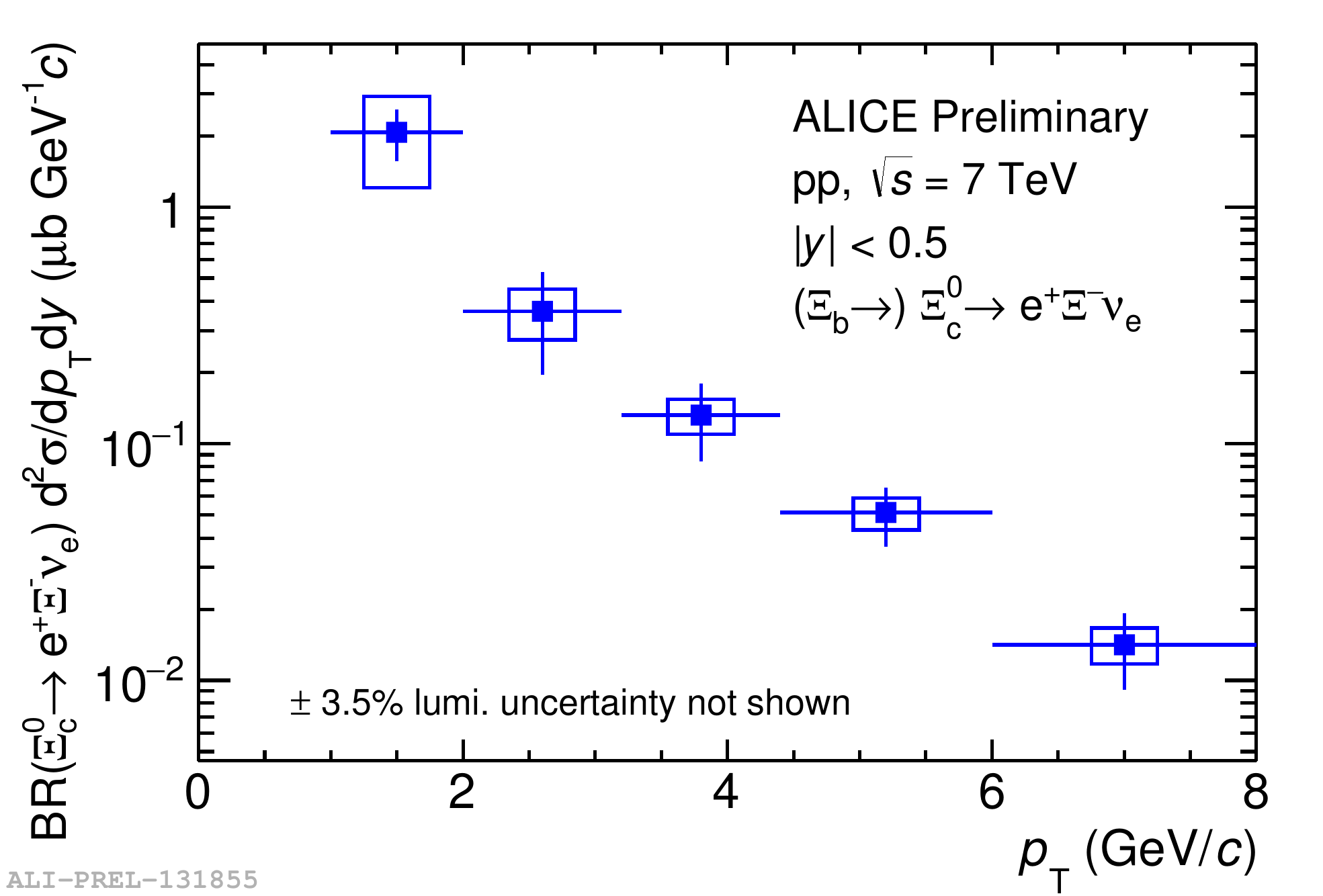}
	\includegraphics[width=6cm,clip]{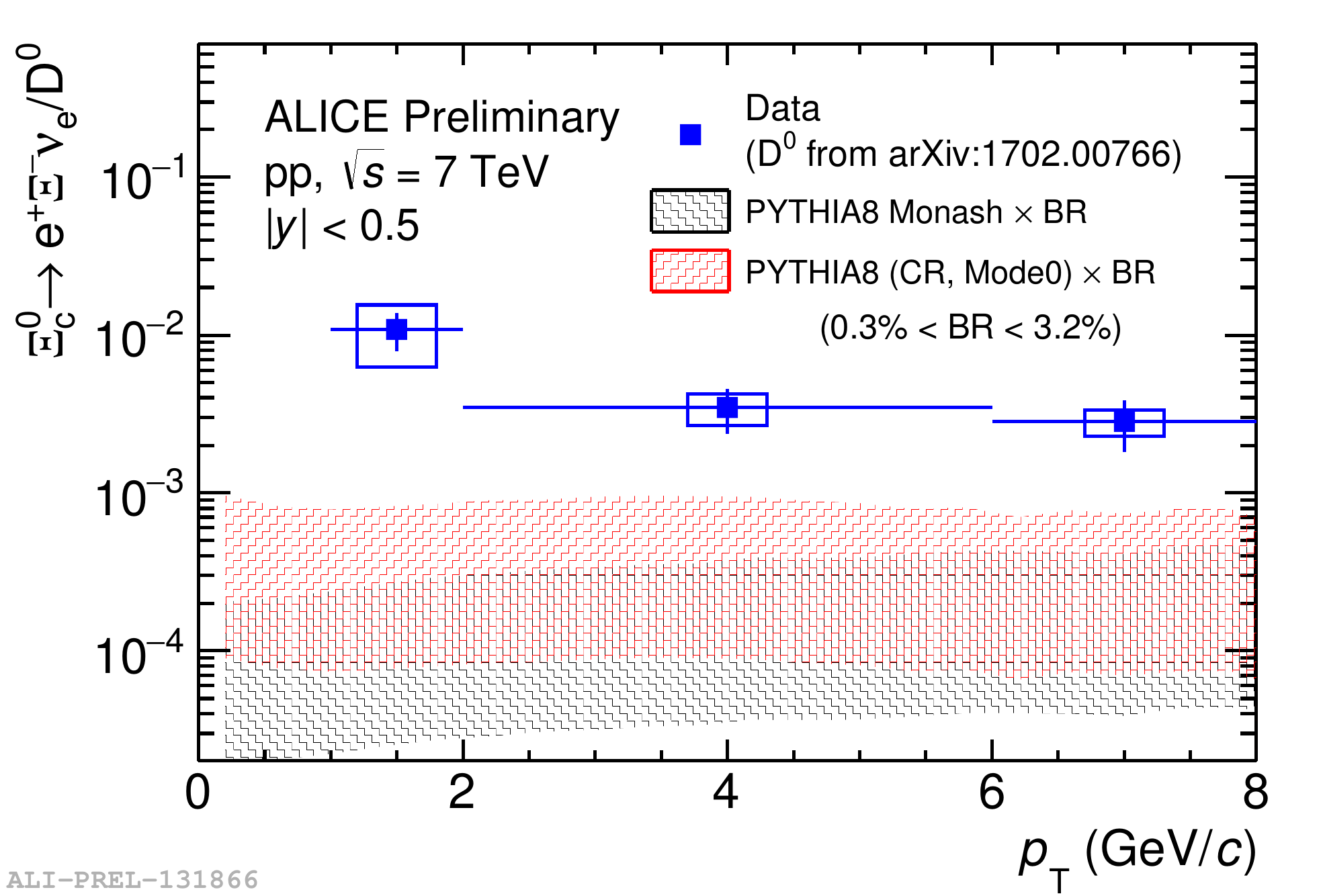}
	\caption{Left: The $\pt$-differential cross section times branching ratio of the $\Xic$ baryon in $\pp$ collisions at $\sqrtS = 7 ~\tev$. Right: The $\XictoenuXi / \Dzero$ ratio compared to theoretical predictions.}
	\label{fig:Xic}       % Give a unique label
\end{figure}

Figure \ref{fig:Xic} shows the $\pt$-differential cross section times branching ratio of the $\Xic$ baryon (including prompt and non-prompt contributions), and the baryon-to-meson ratio $\XictoenuXi / \Dzero$ in comparison to predictions from {\sc pythia} with the aforementioned tunes. The shaded band for the models spans the range of theoretical predictions for the $\XictoenuXi$ branching ratio \cite{PerezMarcial:1989yh,Singleton:1990ye,Cheng:1995fe}. As for the $\Lcplus / \Dzero$ ratio, all predictions significantly underestimate the data.

\section{Summary}

Recent charmed meson and baryon measurements by the ALICE collaboration have been presented.
The $\Rppb$ of D mesons is found to be consistent with unity. The $\Qcp$ of $\Dzero$ mesons shows a hint of $\Dzero$ enhancement at $3 < \pt < 8 ~\gevc$ in central $\pPb$ collisions. The cross section of the $\Lcplus$ baryon and the $\Lcplus / \Dzero$ and $\XictoenuXi / \Dzero$ ratios are found to be underpredicted by theoretical calculations. Finally the $\Rppb$ of $\Lcplus$ baryons is found to be consistent with unity, with the D-meson $\Rppb$ and with theoretical predictions within current uncertainties.

 \bibliography{my-bib-database.bib} 
%\begin{thebibliography}{}
%Journal Author, Journal \textbf{Volume}, page numbers (year)
%Book Author, \textit{Book title} (Publisher, place, year) page numbers
% etc
%\end{thebibliography}

\end{document}